\documentclass[prx,aps,secnumarabic,amsmath,amssymb,twocolumn,notitlepage,longbibliography]{revtex4-2}
\usepackage[colorlinks=true,urlcolor=blue]{hyperref}
\usepackage[normalem]{ulem}
\usepackage[utf8]{inputenc}
\usepackage[english]{babel}
\usepackage[T1]{fontenc}
\usepackage{mathtools}
\usepackage{mathrsfs}
\usepackage{amsfonts}
\usepackage{bbm}
\usepackage{amsthm}
\usepackage{xcolor}
\usepackage{bbold}
\usepackage{bm}
\usepackage{soul}
\usepackage{balance}

\definecolor{red}{rgb}{0.75,0,0}
\definecolor{blue}{rgb}{0,0,0.75}
\definecolor{green}{rgb}{0,0.5,0}

\newcommand{\Tr}{\operatorname{Tr}}%

\begin{document}

\title{Topological Phases and Curvature-Driven Pattern Formation in Cholesteric Shells}
\author{G. Negro$^1$}
\author{L.N. Carenza$^2$}
\author{G. Gonnella$^3$}
\author{D. Marenduzzo$^1$}
\author{E. Orlandini$^4$}
\email{carenza@lorentz.leidenuniv.nl}
\affiliation{$^1$ SUPA, School of Physics and Astronomy, University of Edinburgh, Peter Guthrie Tait Road, Edinburgh, EH9 3FD, UK}
\affiliation{$^2$ Instituut-Lorentz, Universiteit Leiden, P.O. Box 9506, 2300 RA Leiden, The Netherlands}
\affiliation{$^3$ Dipartimento di Fisica, Università degli Studi di Bari and INFN, Sezione di Bari, via Amendola 173, Bari, I-70126, Italy.}
\affiliation{$^4$ Dipartimento di Fisica e Astronomia, Università di Padova, 35131 Padova, Italy}
\date{\today}

\begin{abstract}
We study the phase behaviour of cholesteric liquid crystal shells with different geometries. We compare the cases of tangential and no anchoring at the surface, focussing on the former case, which leads to a competition between the intrinsic tendency of the cholesteric to twist and the anchoring free energy which suppresses it.We then characterise the topological phases arising close to the isotropic-cholesteric transition.These typically consist of quasi-crystalline or amorphous tessellations of the surface by half-skyrmions, which are stable at lower and larger shell size respectively. For ellipsoidal shells, defects in the tesselation couple to local curvature, and according to the shell size they either migrate to the poles or distribute uniformly on the surface. For toroidal shells, the variations in the local curvature of the surface stabilises heterogeneous phases where cholesteric or isotropic patterns coexist with hexagonal lattices of half-skyrmions.
\end{abstract}

\maketitle

\section{Introduction}

Cholesteric liquid crystals (CLC) are fascinating soft materials that naturally exhibit a wide zoology of topological phases and singularities --the so-called topological defects--and therefore they provide a prolific ground for testing abstract topological theories in many areas of physics and mathematics~\cite{Wright89,science.1201968,Hindmarsh_1995,science.1134796,C3SM50605C}. 
The origin of the interest of the scientific community however, is due to various striking optical properties of CLC. For instance, they have a chiral pitch of the same order of magnitude as the wavelength of visible light, respond to applied electromagnetic fields and exhibit important birefringence properties. Therefore they are extremely valuable in a wealth of applications to optical devices, liquid crystal-based lasers and nanotechnology~\cite{bluephaselasers,skyrmionssmalyukh,hopfionsmalyukh,Schwartz2018}, as well as in bio-inspired realizations~\cite{carenza2019,physicaA2019,cates_chirality}.

Among others, Blue Phases (BP) are examples of CLC that occur in proximity of the isotropic-nematic transition and manifest themselves as a network of double twist cylinders that self-assemble into three-dimensional (3D) structures~\cite{Wright89,Alexander2006,alexander2009,cates_henrich}, which appear in vivid shades of blue -- to which they owe their name. 
In 3D BPs come in three different forms. BPI and BPII feature cubic arrangements in the network of disclination lines and therefore have a crystalline nature, and find industrial applications in designing fast light modulators~\cite{peng_15}, lasers~\cite{bluephaselasers}, display devices~\cite{kikuchibpdevice}  and tunable photonic crystals~\cite{oton_20}.
Instead, BPIII still features a network of disclination lines~\cite{Henrich2011}, but arranged in a disordered fashion to yield an amorphous structure~\cite{Henrich2011,Gandhi2017}, and it is also referred to as the \emph{blue fog}, with the name deriving from a loss in the observed brightness of the coloured texture.

CLC --and BPs in particular-- are fully 3D structures and, when confined within restricted regions, competition between geometrical constraints and 3D defects gives rise to a phenomenon known as topological frustration that may produce a plethora of new defect arrangements or topological phases~\cite{depablo2015,noi,Teresa2017,Yang2022,depablo2015}.
Within quasi-2D CLC samples, the chirality of the molecules favours the onset of double-twisted cylinders, which can be viewed as 
topological quasi-particles known as half-skyrmions (or merons). In other quasi-2D geometries, even more exotic structures such as knotted and linked disclination loops (hopfions) can be observed~\cite{julia2019,PhysRevE_97_062706,hopfionsmalyukh}.

The emerging physics becomes even richer when the CLC are confined in a region bordered  by curved surfaces. Such cases can be experimentally obtained  by encapsulating a small amount of liquid crystal within spherical shells~\cite{rassegnadifettiliquidcrystalsgocce}.
Confined nematic LC~\cite{Kamien2002,D1SM00719J} yield a panoply of different configurations were observed depending on the alignment (homeotropic or tangential) and thickness of the liquid crystalline shell. For instance, in the case of tangential anchoring, topological confinement follows from the Gauss-Bonnet theorem, stating that the total topological charge (the sum of the charge of all defects) must equal the Euler characteristic of the bounding geometry. Therefore, for spherical shells this requires a total $s=+2$ defective charge which can be realised either with four $+1/2$ disclinations, or with two boojums, each carrying $+1$ topological charge, or even with more complex states~\cite{rassegnadifettiliquidcrystalsgocce}. 
Very recently this technique has been extended to CLC~\cite{Teresa2017} which lead to the additional appearance of skyrmions, merons and other transitory states, thanks to the competition between the anchoring direction of the LC molecules at the boundary, the surface curvature and the cholesteric pitch of the LC, which introduces yet an additional length-scale.

Notwithstanding this initial experimental evidence, little is known from a theoretical perspective on the ordering properties of CLC confined in non-flat geometries. To fill this gap, in a recent paper~\cite{noi} we made use of the Landau-de Gennes field theory to study the equilibrium properties of CLC confined into shells (cholesteric shells).
By means of lattice Boltzmann simulations, we found that cholesteric shells support various types of topological phases that are  completely absent on the quasi-2D flat geometries, analysed for instance in~\cite{Metselaar2018}.

In particular, under strong geometrical confinement obtained by placing the CLC on a shell with a radius comparable with the cholesteric pitch, quasi-crystalline lattices of half-skyrmions were observed with several possible mixtures of polygonal tessellations, depending both on the radius and the chirality of the CLC. Interestingly, yet another phase appeared at larger chirality, with the loss of quasi-crystalline order leading to an amorphous patching of half-skyrmions. This may be viewed as a 2D counterpart of the 3D amorphous blue phase, the blue fog. 
These results are especially notable if one considers that this wide zoology of possible states were observed in absence of anchoring of the LC molecules on the shell surface. Anchoring is indeed expected to play a significant role on the stability and degree of ordering of the topological phases discovered in~\cite{noi}: for instance, a finite anchoring on a thin shell it should suppress twisting of the director field, which is thermodynamically favoured in CLC. Moreover, studying the stability of topological phases in presence of perturbing interactions, as anchoring, is relevant even from an experimental perspective as liquid crystalline shells can be stabilized in the lab through surfactants which often favour a preferential orientation of liquid crystal molecules at the interface~\cite{Bukusoglu2016,Darmon2016}.
Another potentially important issue which is poorly investigated is the effect of curvature on the half-skyrmion organisation and on topological phases in general. This is a relevant issue to explore given that soft shells can easily be deformed locally, thereby providing a way to control local curvature experimentally. 

In this article we will address these two open questions. First, we will consider the case of a CLC spherical shell with tangential anchoring at the surface and we will show how an orientational energetic penalty can significantly stabilize novel quasi-crystalline structures by hindering the proliferation of defects in the tessellation. Furthermore, we will show how geometries with non-uniform Gaussian curvature can be exploited to control defect positions, by considering both ellipsoidal and toroidal shells.

The article is organized as follows. In Section~\ref{sec:model} we provide details of Landau-de Gennes free-energy of cholesteric shells with tangential anchoring and their associated dynamical equations.
In Section~\ref{sec:phasediagram} we determine the equilibrium phase diagram of the system  and characterize the order and regularity of the discovered topological phases in terms of both a generalised hexatic order parameter and the Bessel structure factor. To pinpoint the role of the anchoring we compare these results with those previously obtained in absence of LC anchoring at the surface~\cite{noi}.
In Section~\ref{sec:curvature} we extend our analysis to ellipsoidal and toroidal cholesteric shells. The  goal here is to understand how the presence of a non-homogeneous local curvature impacts on the stability and the patterning  of the observed structures.
We will finally discuss the results in Section~\ref{sec:conclusions}.

\section{Model}
\label{sec:model}
We start by introducing the relevant order parameters for describing the state of the system. The orientational properties of the liquid crystal (LC) are customarily described by the symmetric and traceless tensor $\bm{Q}(\bm{r})$. The principle direction of the  tensor, defined as the eigenvector $\bm{n}$ of the largest eigenvalue, represents the local alignment direction of the liquid crystal molecules in a certain point in space, while the amplitude $\sqrt{\Tr(\bm{Q}^2)}$ is a measure of the degree of liquid crystalline order. In the following we will identify topological defects on the shell surface by measuring the degree of biaxiality of the LC, looking at the second parameter of the Westin metrics $c_p$\cite{callan2006} \footnote{The biaxiality parameter is defined as $c_p = 2(\tilde\lambda_2-\tilde\lambda_3)$, where $\tilde\lambda_1$, $\tilde\lambda_2$ and $\tilde\lambda_3$ (with $\tilde\lambda_1\ge\tilde\lambda_2\ge\tilde\lambda_3$) are three eigenvalues of the positive definite matrix $G_{\alpha\beta}=Q_{\alpha\beta}+\delta_{\alpha\beta}/3$~\cite{callan2006}.}. 
The phase field, which determines the geometry on which we will confine the liquid crystal, is defined by the scalar field $\phi(\bm{r})$. Finally, the velocity field $\bm{v}$ captures the local velocity of the liquid phase.

\subsection{Free Energy}
The ground state of the system is defined by the free energy
\begin{eqnarray}
F^{Q}&=& \int d{\bf r} \left \{ A_0 \left [ \dfrac{1}{2}  \left(1 - \dfrac{\chi}{3} \right)\bm{Q}^2 -  \dfrac{\chi}{3} \bm{Q}^3 +  \dfrac{\chi}{4} \bm{Q}^4\right ]\right. \nonumber \\
&+& \frac{L}{2} \left [ (\nabla\cdot \bm{Q})^2  + \left ( \nabla \times \bm{Q} + 2q_0 \bm Q)^2 \right ] \right \}.
\label{eqn:freeEchol}
\end{eqnarray}
Here, the terms on the first line, proportional to the bulk constant $A_0$, capture the first-order isotropic-nematic transition. This occurs when the temperature-like parameter $\chi>\chi_{cr}=2.7$.
The terms on the second line, proportional to the elastic constant $L$, are the elastic contributions to the free energy and account for the elastic energy penalty due to LC deformation. 
For non-zero values of the parameter $q_0$, the mirror symmetry is broken and chiral states emerge at equilibrium. In particular, in the following, we will consider the case of right-handed twist ($q_0>0$) with the equilibrium pitch given by $p_0=2\pi /q_0$ in bulk systems.
It is worth noticing that the saddle splay term of the director field theory is inherently effectively incorporated in our model, as shown  in\cite{Kos2016}.

The phase field ground state is determined by minimising the following free energy:
$$
F^{\phi}=\int d{\bf r} \left [  \dfrac{a}{4} \phi^2 (\phi-\phi_0)^2 + \dfrac{k_\phi}{2} (\nabla \phi)^2 \right].
\label{eqn:freeEphi}
$$
For $a>0$, the polynomial term has two possible equilibrium values (minima) at $\phi=0,\phi_0$.
The two material parameters $a$ and $k_\phi$ account for the surface tension $\gamma = \sqrt{8 a k_\phi/9}$ and the interface width $\xi_\phi = \sqrt{2 k_\phi/a}$.

To create and stabilise a LC shell we require the temperature-like parameter $\chi$ to depend on the gradients of the phase field $\phi$ as follows
$$
\chi=\chi_0 + \chi_s (\nabla \phi)^2.
$$
In this way the Q-tensor attains non-zero values only in those regions where $|\nabla \phi|>\sqrt{(\chi_{cr}-\chi_0)/\chi_s}$ (see Appendix \ref{sec:appendix}). 
For spherical droplets, $\phi$ is $\phi_0$ inside the droplet and $0$ otherwise, so that the square of the gradient $(\nabla \phi)^2$ is maximal at the interface, and approaches $0$ sufficiently far from it. In our model $\chi$ depends on  $(\nabla \phi)^2$ in such a way that the system is liquid crystalline at the interface and isotropic away from it, resulting in a thin shell of liquid crystal with isotropic fluid inside and outside.
Different (non-spherical) shell geometries can be obtained by letting $\phi_0$ attain a spatial modulation:
\begin{equation}
\phi_0(\bm{r})= \begin{cases} 
\bar{\phi_0} \text{ if }  \mathcal{T}(\bm{r})<0\\
0 \text{ otherwise }
\end{cases} 
\label{eqn:parametric_non_spherical}
\end{equation}
where $ \bar{\phi_0}$ is a constant and $\mathcal{T}(\bm{r})$ is the parametric equation of a generic manifold. For instance, in the case of a torus with radii $R_1$ and $R_2$ (with $R_1>R_2$) $ \mathcal{T}(x,y,z)=(R_1-(x^2+y^2)^{1/2})^2 +z^2-R_2^2$. Therefore, the interface is formed where $\mathcal{T}\approx 0$, hence on the torus surface. Analogously, for the case of an ellipsoid $ \mathcal{T}(x,y,z)= x^2/R_{1}^2 + y^2/R_{2}^2 + z^2/R_{3}^2 - 1$, being $R_1,R_2,R_3$ the three semiaxis of the ellipsoid. This approach is appropriate for cases where we are not interested in deformations or motion of the shell, such as ours.
Notice that the phase field free energy in Eq.~\eqref{eqn:freeEphi}, gives a description of the interface equivalent to that of the Helfrich model (see \emph{e.g.}~\cite{Liu2003}). 
The gradient square term in the coupling with the liquid crystal is equivalent to a delta function selecting points at the interface. This serves as a regularization in the context of phase field models~\cite{Lee2012}. 

Finally, soft anchoring at the surface of the shell is imposed through an additional coupling term in the free energy involving the Q-tensor and the gradient of the phase field,
\begin{equation}
{F}^{anch}= \int d{\bf r}\beta \left[ \nabla \phi \cdot  \bm{Q} \cdot \nabla \phi \right].
\end{equation}  
Tangential anchoring is achieved by choosing $\beta>0$: we focus on such a case in this work.

\subsection{Dimensionless numbers}
The dimensionless numbers determining the behaviour of the system are as follows: \emph{(i)} the reduced temperature $\tau={9(3-\chi)}/{\chi}$, \emph{(ii)} the chirality strength $\kappa=\sqrt{108 q_0^2 L /(A_0 \chi)}$ proportional to the ratio between cholesteric pitch and nematic coherence length $\xi_n=\sqrt{L/A_0}$~\cite{alexander2009,Wright89}, and \emph{(iii)} the ratio between shell radius and cholesteric pitch, $R/p_0$~\cite{depablo2015}.Moreover, we choose parameters so that the interface thickness $\xi_\phi \ll p_0$.
In fully 3D systems and low chirality ($\kappa \ll 0.5$) for $\tau<\tau_c(\kappa)=\frac{1}{2} \left[1-4 \kappa^2 + \left(1+ 4 \kappa^2/3\right)^{3/2}  \right]$, Eq.~\eqref{eqn:freeEchol} is minimised by the helical phase~\cite{brazovskii1975phase} that manifests itself as unidirectional twisting of liquid crystalline molecules and  with equilibrium pitch $p_0=2\pi q_0^{-1}$. As chirality grows larger ($\kappa \gg 0.5$), double twist of the cholesteric helix becomes favorable and a new metastable phase--the blue phase--appears. In the mean-field approach of Ref.~\cite{brazovskii1975phase}, for large values of the reduced chirality $\kappa$, the system is isotropic when $\tau > 0.8 \tau_c({\kappa=0})$, it develops blue phases when the reduced temperature is comprised between $0.8 \tau_c({\kappa=0})$ and $\tau_c(\kappa)$, and is in its helical phase for $\tau<\tau_c(\kappa)$.
In 2D, blue phases appear as a regular half-skyrmion lattice with hexagonal symmetry. A numerical phase diagram of the stability of such phases in 2D is provided in~\cite{Metselaar2018,julia2019}.

\subsection{Dynamical Equations}
Nemato-hydrodynamics is ruled by the incompressible Navier-Stokes equation for the flow field $\bm{v}$ and the Beris-Edwards equation for the Q-tensor.
The former is given by
$$
\rho(\partial_t \bm{v} + \bm{v} \cdot \nabla \bm{v}) = -\nabla p + \nabla \cdot ( \bm{\sigma}^{visc} + \bm{\sigma}^{el}),
$$
where $\rho$ is the total (constant) density of the fluid and $p$ is the hydrodynamic pressure. The viscous contribution to the stress tensor is given by
$$\sigma^{visc}_{\alpha \beta}=\eta (\partial_\alpha v_\beta + \partial_\beta v_\alpha)$$ 
where $\eta$ is the nominal viscosity of the isotropic fluid. 
The elastic stress
\begin{equation}
\begin{split}
\sigma_{\alpha \beta}^{el} = & -\tilde{\xi} H_{\alpha \gamma} \left(Q_{\gamma \beta} + \dfrac{1}{3} \delta_{\gamma \beta} \right) -\tilde{\xi}  \left(Q_{\alpha \gamma} + \dfrac{1}{3} \delta_{\alpha \gamma} \right) H_{\gamma \beta} \\  & + 2\tilde{\xi} \left(Q_{\alpha \beta} - \dfrac{1}{3} \delta_{\alpha \beta} \right) Q_{\gamma \mu} H_{\gamma \mu} + Q_{\alpha \gamma} H_{\gamma \beta}  - H_{ \alpha \gamma} Q_{\gamma \beta} \\ & -\partial_{\alpha}Q_{\gamma\nu}\frac{\partial f}{\partial\partial_{\beta}Q_{\gamma\nu}}
\end{split}
\end{equation} 
is responsible for the backflow that originates from deformations in the LC arrangement. [In the equation above, $f$ denotes the free energy density corresponding to the free energy $F^{Q}+F^{\phi}+F^{anch}$.] Here, $\bm{H}=-\frac{\delta F}{\delta \bm{Q}}+ \frac{\bm{I}}{3} Tr \left(\frac{\delta F}{\delta \bm{Q}} \right)$ is the molecular field (with ${F}={F}^{Q}+ {F}^{\phi} + {F}^{anch}$ the total free energy), $f$ is the free energy density corresponding to $F$,  while $\tilde{\xi}$ is the flow-alignment parameter which controls the aspect-ratio of the liquid crystal molecules and aligning properties to the flow (we chose $\tilde{\xi}=0.7$ to consider flow-aligning rod-like molecules).

\begin{figure*}[ht!]
\centering\includegraphics[width=1.0\textwidth]{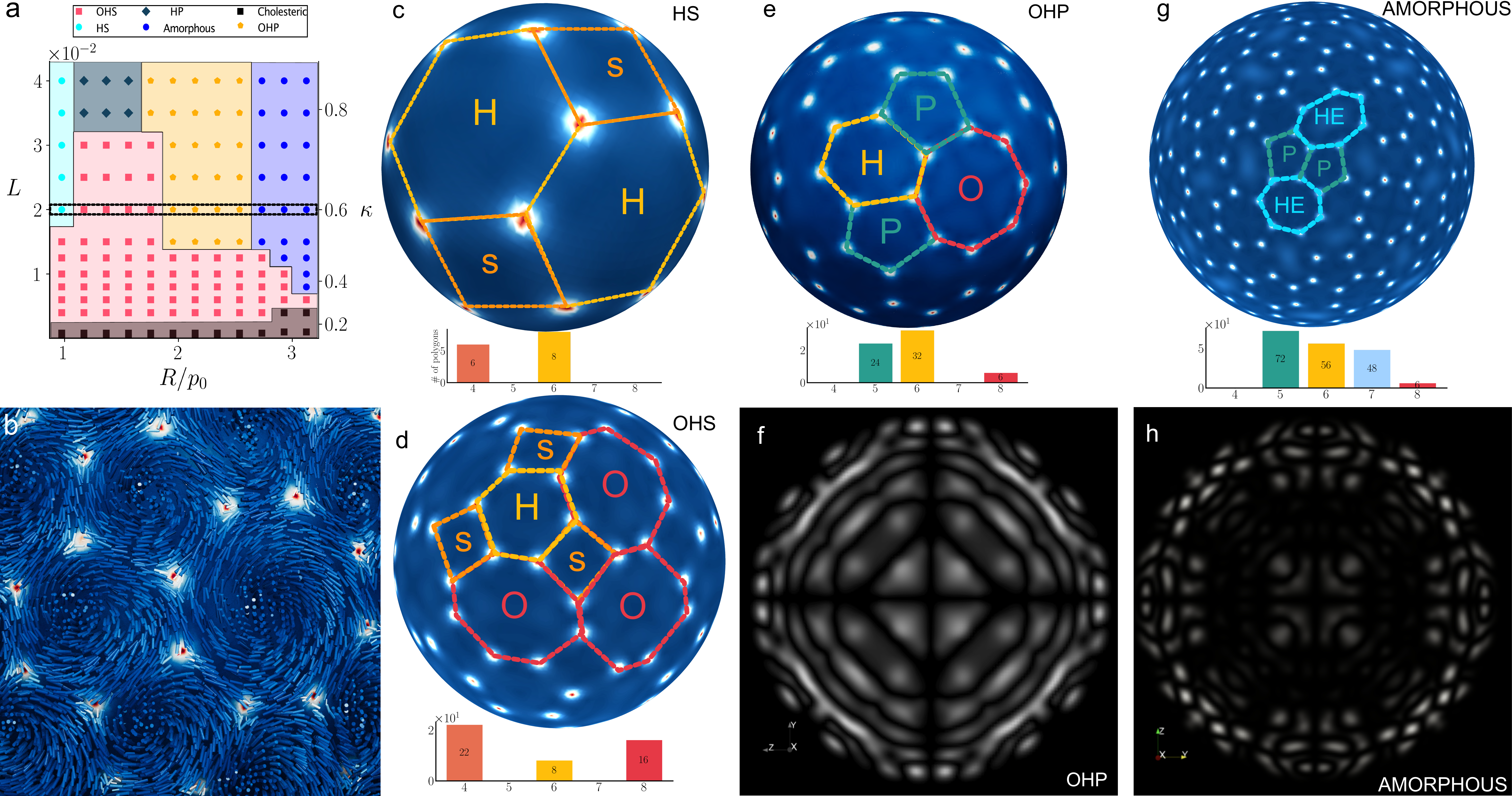}
\caption{\textbf{Phase behaviour and typical shell configurations with tangential anchoring.} Panel~(a) shows the phase diagram in the $\kappa-R/p_0$ plane for spherical shells with tangential anchoring ($\beta=8 \times 10^{-3}$). 
In the legend, $S$=square, $P$=pentagons, $H$=hexagons, $O$=octagons, $HE$=heptagons.
Panel~(b) shows the pattern of director field of half-skyrmions and  $-1/2$ defects defining octagons, hexagons and pentagons for a magnified region of the configuration in panel~(d).
Three quasi-crystalline configurations are shown in panels~(c-e), for the same values of reduced chirality $\kappa=0.6$ (or equivalently $L=0.02$) and varying values of the radius $R$. These are: (c) $HS$ at $R/p_0=1$ ($R=25$), (d) $OHS$ at $R/p_0=1.2$ ($R=30$) and (e) $OHP$ at $R/p_0=2$ ($R=50$).  
Panel~(g) shows an amorphous configuration at $R/p_0=3.2$ ($R=80$), for the same value of reduced chirality $\kappa=0.6$ as previous cases. 
The dashed line in panel (a) highlights the region with constant $\kappa$ where the configurations (c-f) were taken. 
The color code in panel~(b-f) corresponds to the isotropic parameter $c_s$ of the Westin metrix~\cite{callan2006}: red-white regions define defect positions ($\bm{Q} \sim 0$), while blue ones are ordered.
Panels~(f) and (h) show the Schlieren textures corresponding to the $OHP$ and amorphous configurations, respectively. The stability of the observed configurations has been tested on several runs for a selected number of parameter values by varying the random initial conditions and fixing the physical parameters of the simulation.}
\label{fig1}
\end{figure*}
The Beris-Edwards equation for $Q$-tensor evolution is given by
$$
(\partial_t + \bm{v}\cdot \nabla)\bm{Q} - \mathcal{S}(\nabla \bm{v}, \bm{Q}) = \Gamma \bm{H},
$$
where $\Gamma$ is the rotational viscosity which measures the relative importance of advective interactions with respect of the driving effect of 
the molecular field $\bm{H}$ and
\begin{equation}
\begin{split}
\mathcal{S}(\nabla \bm{v}, \bm{Q})&=(\tilde{\xi} \mathbf{D} + \bm{\Omega})(\mathbf{Q}+\mathbf{I}/3) + (\mathbf{Q}+\mathbf{I}/3)(\tilde{\xi} \mathbf{D}  - \bm{\Omega})\\ &- 2 \tilde{\xi} (\mathbf{Q}+\mathbf{I}/3) Tr (\mathbf{Q}\mathbf{W})
\end{split}
\end{equation}
is the strain-rotational derivative. Here, $\bm{D}$ is the strain-rate tensor and $\bm{\Omega}$ the vorticity tensor, respectively given by the symmetric and anti-symmetric part of the velocity gradient tensor $\bm{W}=\nabla \bm{v}= \partial_\beta v_\alpha$. 

Finally, in this paper we are not interested in the formation process of spherical shells but rather on the relaxation of the liquid crystal on curved geometries. This is equivalent to requiring that the typical relaxation time of the liquid crystal dynamics are long compared with those of the phase field.
Therefore, after thermalization, we assume the phase field $\phi$ as a static non-evolving field and its configuration is obtained by a free energy energy minimisation procedure for each geometry considered.

\section{Spherical shells: equilibrium phase diagram and anchoring effects}
\label{sec:phasediagram}
We start our discussion by considering spherical shells. We fix the reduced temperature $\tau=0.540$ and the cholesteric pitch $p_0=25.64$. We enforce soft tangential anchoring by setting the anchoring constant $\beta = 8 \times 10^{-3}$ and look for the equilibrium states obtained by varying values of the chirality strength $\kappa$ and of the ratio between the shell radius and the helix pitch, $R/p_0$. These are in turn tuned by varying respectively the elastic constant $L$ and, for spherical shells, their radius  $R$.

In Fig.~\ref{fig1}(a) we report the numerical equilibrium phase diagram partitioned according to different topological phases obtained by varying the chirality and the size of the confining geoemetry.
In addition to the expected cholesteric and isotropic phases (not shown) observed respectively for small ($\kappa \lesssim 0.2)$ and very large values ($\kappa \gtrsim 1.0$) of chirality, other stable phases appear for intermediate values of $\kappa$ and $R/p_0$. These  topological phases are characterised by different types of defect patterns. More precisely, a network of $-1/2$ point defects is observed, defining  the vertices of a polygonal tessellation of the surface with half-skyrmions positioned at the centres of the polygons--see  the director field in Fig.~\ref{fig1}(b). 
Notice that  hexagonal half-skyrmion lattices have been previously observed in a flat 2D geometry~\cite{selinger,Metselaar2018}. However, since $n$-edges polygon in the tessellation contributes to a topological charge of $1-n/6$, a pure hexagonal lattice of defects is forbidden on a spherical curved geometry by the Gauss-Bonnet theorem~\cite{Kamien2002,rassegnadifettiliquidcrystalsgocce}, as this configuration would result into a vanishing total topological charge, rather than the required value of $+2$.
As a consequence, the polygonal tessellation that we observe on spherical shells  comprises polygons other than hexagons. 
Some examples of configurations are reported in panels (c-f) obtained for $L=2\times 10^{-2}$ and  four increasing values of $R$.

More precisely, the topological constraint provided by the Gauss-Bonnet theorem for the sphere, once expressed through the Euler formula for polyhedra, reads $\sum \left (1 -\frac{n}{6}\right)N_n=2$, and provides a condition on the number of $n$-edges polygons $N_n$.

According to the estimated phase diagram of Fig.~\ref{fig1}(a),  if we fix for instance $\kappa=0.6$, one can notice that in the range $1 \le R/p_0 \le 2.6$  the defect structures are locally regular with various possible tessellations that  differ by the number and types of polygons, as $R/p_0$ is varied.
More specifically, at $R /p_0 =1$  we observe a phase characterised by a regular network of $8$ hexagons and $6$ squares which we  denote by $HS$, see Fig.~\ref{fig1}(c).  As $R /p_0$ is increased, a new defect structure is observed, where also few octagons appears in the tessellation. In this $OHS$ phase, each polygon is in contact with  an equal number of polygons of different types disposed in an ordered fashion: for instance a hexagon is surrounded by three octagons alternated by squares, see Fig.~\ref{fig1}(d). 
For  $R /p_0>1.8$ the number of defects further  increases and, in order to satisfy the Gauss-Bonnet theorem, a new  defect pattern, where the squares of the $OHS$ phase are replaced by pentagons, emerges.  This regular mixture of octagons, hexagons and pentagons is denoted by $OHP$, see \ref{fig1}(e,f). 
Note that all these three phases display a local order that is reminiscent of that of quasicrystals, \emph{i.e.} structures which lack the typical discrete translational symmetries of periodic crystals but display discrete spatial Fourier spectra (see next subsection). 
The quasi-crystalline character of the configurations is however lost if $R /p_0 > 2.6$. In this regime the number of defects, that grows with the system size, largely exceeds the minimal one required by the spherical topology and this overabundance of defects appears as pentagons-heptagons pairs of opposite topological charge (see Fig.~\ref{fig1}g,h). 
Moreover the neighbouring structure of the  polygonal tessellation is now disordered: no simple correlation between the types of neighbouring polygons is observed and  the resulting half-skyrmion arrangement is akin to an amorphous lattice (see Section~\ref{sec:struct_fact}).

\begin{figure}[t]
\centering\includegraphics[width=0.9\columnwidth]{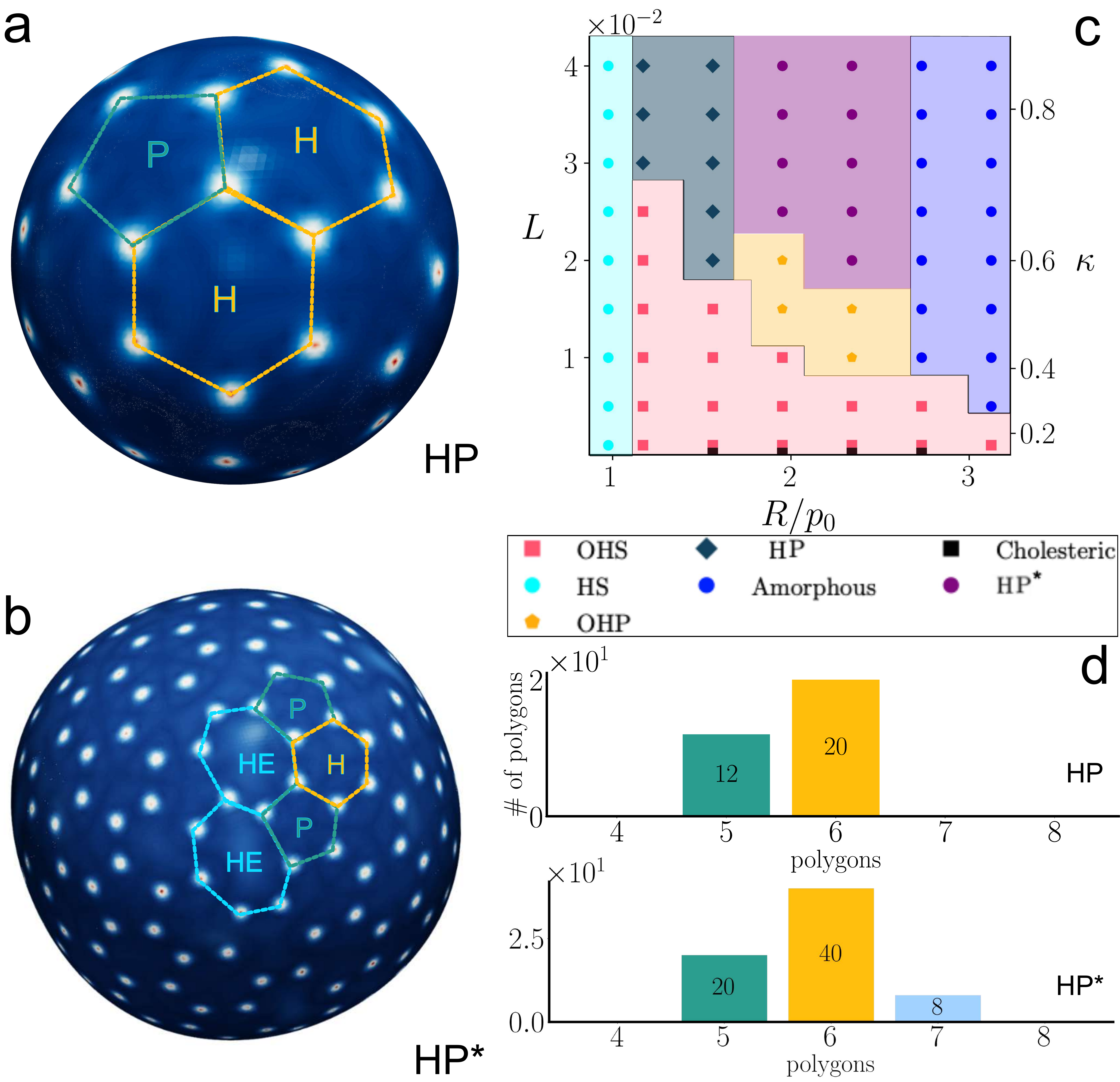}
\caption{\textbf{Spherical shells with free anchoring.} 
\textbf{(a-b)} Comparison between the $HP$ \emph{fqc} state at $R=40,\  L=3.5\times 10^{-2}$, with tangential anchoring (panel a), and the  $HP^*$ state at $R=60, \ L=2.0\times 10^{-2}$, for spherical geometry with free anchoring (panel b).  The color code is the same as that of Fig.~\ref{fig1}(b-e). \textbf{(c)} Phase diagram in the $\kappa-R/p_0$ plane, for spherical shells with free anchoring. 
\textbf{(d)} Histogram of polygonal occurrences for the $HP$ (top) and $HP^*$ (bottom) configurations. The figure is a summary of previous results presented in Ref.~\cite{noi}.
}
\label{fig2}
\end{figure}

Note that, if we stay within a given topological phase, for instance by keeping $R/p_0$ fixed, and increase the value of $\kappa$, the number of defects increases and so does the number of polygons in the tessellation. For instance in the $OHS$ phase ($R=30$) if we increase the value of $\kappa$ from $0.4$ to $0.6$ the number of octagons and squares increase respectively from $6$ to $16$, and from $12$  to $22$ while the number of hexagons remains the same (not shown).
This can be rationalised by observing that, by increasing the chirality strength $\kappa$ the equilibrium size of the half-skyrmions decreases while the surface of the shell is kept fixed leaving more room to accommodate more defects. However, when the organisation of newly formed  defects cannot satisfy anymore the topological constraint of the sphere  using  the polygon types  of that phase,  the system eventually crosses over to a different topological phase  whose tessellation comprises different types of polygons. 
For instance, at $R/p_0=2$ and $\kappa$ increasing from $0.50$ to $0.52$, the system undergoes a transition from the $OHS$ to the $OHP$ phase (see Fig.\ref{fig1}a).
For the values of parameters explored here the standard football-like configuration of defects composed by 20 hexagons and 12 pentagons is observed for $ 1.2< R/p_0 < 1.6$ when $\kappa> 0.75$. A typical state in this regime is shown in \ref{fig2}a.  Note that Fig.~\ref{fig2} is devoted to the anchoring-free case, discussed in Ref.~\cite{noi}. By a direct comparison between Fig.~\ref{fig1}a and Fig.~\ref{fig2}c, one can here observe that LC anchoring affects partially the equilibrium phase diagram by changing the range of stability in terms of chirality $\kappa$ and $R/p_0$ at which some topological phases are observed. More precisely, one should notice that: (i) the $HS$, $HP$ and cholesteric phases occur for smaller  values of the cholesteric strength $\kappa$;
(ii) more importantly, if anchoring is neglected, the transition between the $HP$ and the amorphous phases turns out to be  mediated  by the presence of an intermediate topological phase, see Fig.\ref{fig2}b. This latter phase, which we will call $HP^*$ in the following, differs from the  $HP$ one  by having a larger number of defects whose excess of topological charges is managed by replacing  some of the hexagons with pentagon-heptagon pairs of total charge zero (see Fig.\ref{fig2}(d)). In this respect, the $HP^*$ phase is closer to the amorphous one and the fact that it is not observed (at least within the range of parameter values considered) for systems with tangential anchoring suggests that the elastic energy penalty at the shell surface due to anchoring hinders the proliferation of defects and keeps their organisation simpler with respect to the free anchoring case. In general, while anchoring does provide a quantitative change in the stability of the specific topological phases presented in the phase diagrams in Fig.~\ref{fig1}(a) and Fig.~\ref{fig2}(b), we do not report significant variations in the liquid crystal pattern in corresponding phases. More explicitly, the optical properties and the Schlieren textures of a certain state (say a $OHP$ configuration) will not appear different in presence (Fig.~\ref{fig1}(f)) or absence (Supplementary Fig.~S6 in Ref.~\cite{noi}) of anchoring.

\begin{figure}[ht!]
\centering\includegraphics[width=1.\columnwidth]{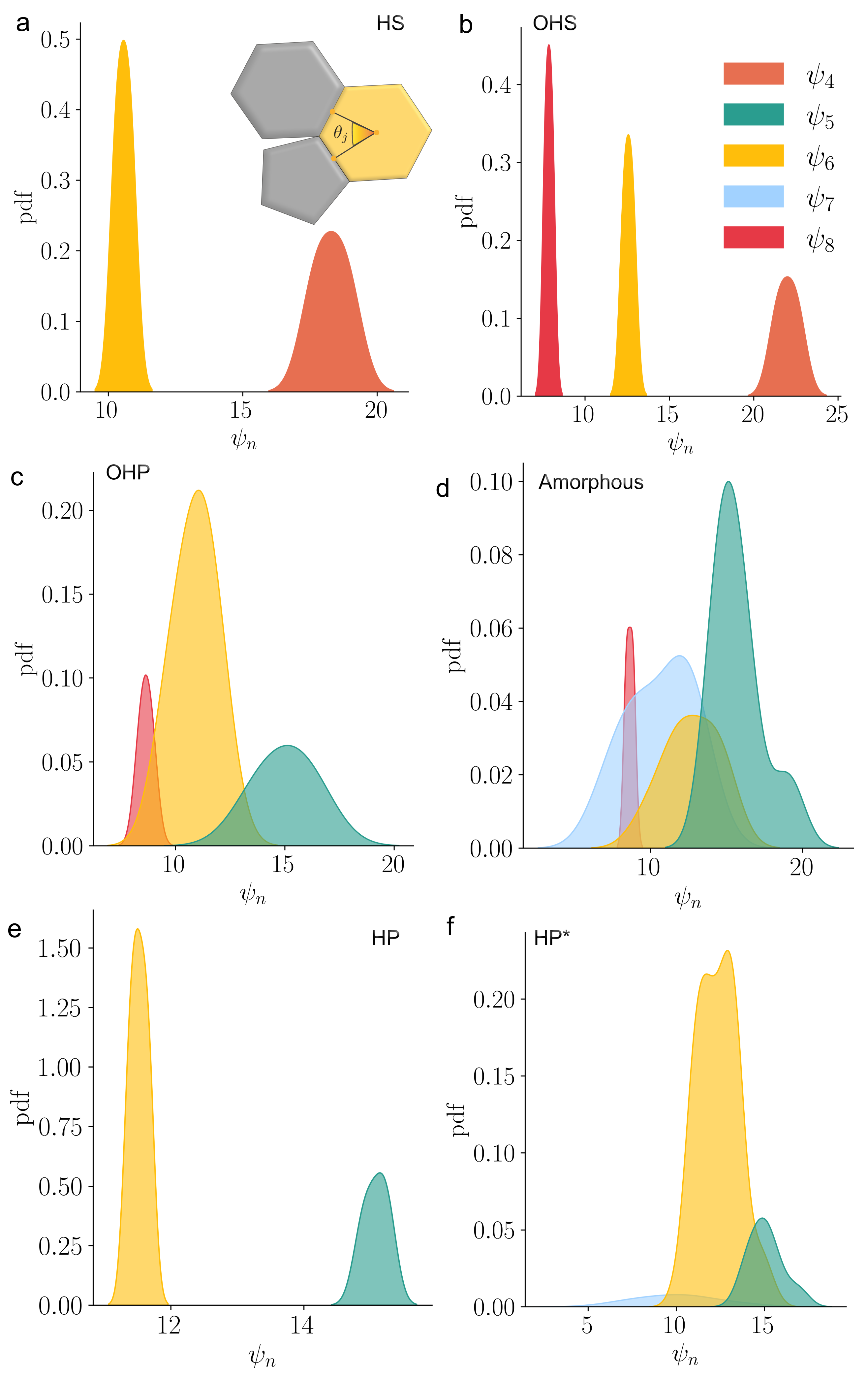}
\caption{\textbf{Order Parameter.} \textbf{(a-f)} Normalized $\psi_n$ \emph{pdf} for the $HS$ (a),  $OHS$ (b), $OHP$ (c), \emph{Amorphous} (d), $HP$ (e)  and $HP^*$ (f) configurations shown in Fig.~\ref{fig1} and Fig.~\ref{fig2}. 	The inset of panel (a) illustrates the definition of $\psi^j_n$.
The expected values for the $HS$ configuration are $\langle \psi_4 \rangle =18.8$, $\langle \psi_6 \rangle = 10.5$; for the $OHS$ $\langle \psi_4 \rangle =22.0, \langle \psi_6 \rangle = 12.6, \langle \psi_8 \rangle = 7.9$; for the $OHP$ $\langle \psi_5 \rangle =16.0, \langle \psi_6 \rangle = 12.2, \langle \psi_8 \rangle = 8.6.$ for the $HP$ $\langle \psi_5 \rangle =15.0, \langle \psi_6 \rangle = 11.5$. Distributions have been normalized such that the sum of the areas under all curves in each panel sums up to $1$.}
\label{fig3}
\end{figure}

\subsection{Order Parameter}
With the aim of quantifying the symmetry of the observed topological phases, we now consider a phenomenological order parameter $\psi_n$, previously introduced in~\cite{noi}. Here the subscript $n$ refers to the component of $n$-edge polygons in the tessellation. The order parameter is defined as follows. Referring to the inset of Fig.~\ref{fig3}a, let $\theta^j$ be the angle formed by the midpoints of the $j$-th pair of neighboring edges with the geometrical centre of the corresponding $n$-edged polygon. Denoting with $\mathcal{N}^{j}_{1}$ and $\mathcal{N}^{j}_{2}$ the number of edges of the two bordering polygons \textcolor{black}{(see grey polygons in the inset of Fig.~\ref{fig3}(b))} the order parameter for the $j$-th pair of neighboring edges is defined  as
\begin{equation}
\psi_n = \theta^j (\mathcal{N}^{j}_{1} + \mathcal{N}^{j}_{2}).
\label{eqn:psi_n}
\end{equation}
By computing the order parameter for all the $\mathcal{N}_n = n N_n$ pairs of edges, belonging to polygons with $n$ edges, one obtains a collection of measurements $\lbrace \psi_{n , j} \rbrace_{j= 1, \dots, \mathcal{N}_n}$ which can be used to quantify the degree of regularity of the polygonal tessellation.

Note indeed that, in a regular lattice, $\psi_n$ depends neither on the particular pair of edges considered, since all pairs are indistinguishable, nor on $n$, as $\mathcal{N}^{j}_{1,2}=n$ and $\theta^j=2\pi/n$. Therefore, for a regular lattice the distribution of $\psi_n$ is a Dirac delta functions peaked at $4\pi$. 
For quasi-crystal phases obtained by patching together regular polygons with a different number of edges, the expectation value of $<\psi_n>$ depends on $n$ since in general the sum $\mathcal{N}^{j}_{1} + \mathcal{N}^{j}_{2}$ in Eq.~\eqref{eqn:psi_n} will differ from $2n$ as polygons with different number of edges contribute to the tessellation. Hence the expected distribution is now a Dirac comb --a combination of delta functions with appropriate weights. 
Conversely, in the case of amorphous lattices, the distribution of $\psi_n$ is expected to broaden and the peaks to flatten, as no regularity would be found either at the level of the polygons or of the tessellation.
Therefore, the information provided by the parameter is twofold: on one hand it captures the regularity of each polygon through the angle $\theta_j$, on the other hand it also determines the regularity of the pattern of neighboring polygons through $\mathcal{N}^{j}_{1}$ and $\mathcal{N}^{j}_{2}$.

\begin{figure}[ht!]
\centering\includegraphics[width=0.95\columnwidth]{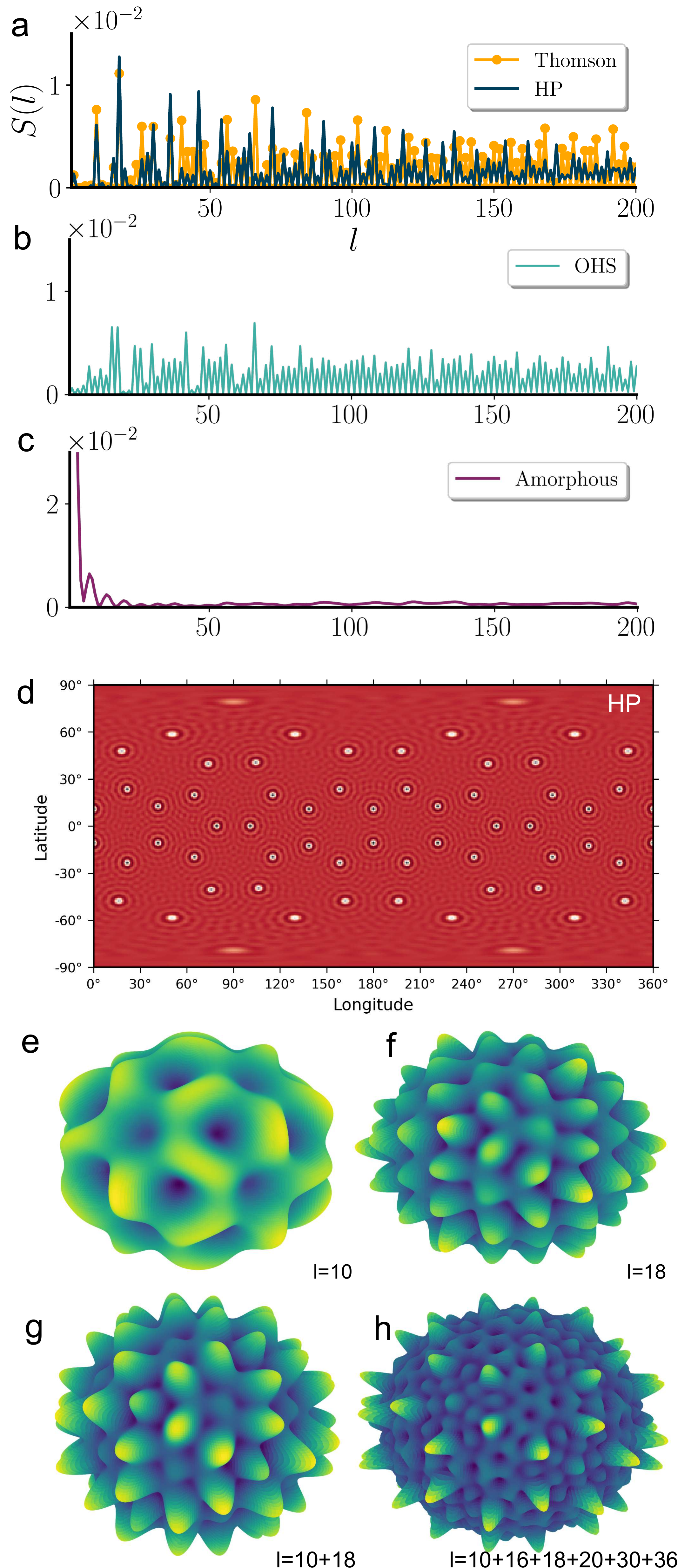}
\caption{ \textbf{Structure Factor.} Panels (a-c) show the  Bessel structure factor of: (i) the  football-like configuration ($HP$, $R=40,L=0.03$),  compared with the exact solution of Thomson problem with 32 points (panel (a)); (ii) the $OHS$ ($R=40,L=0.01$) configuration (panel (b)); and (iii) the amorphous state (panel (c) for $R=80$ and $L=0.04$) . Panel~(d) shows the spherical harmonic projections of the $HP$ configuration. Panels (e-h) show the  superposition of various harmonics in Bessel expansion for the same configuration.}
\label{fig4}
\end{figure}

\begin{figure*}[ht!]
\centering\includegraphics[width=1.\textwidth]{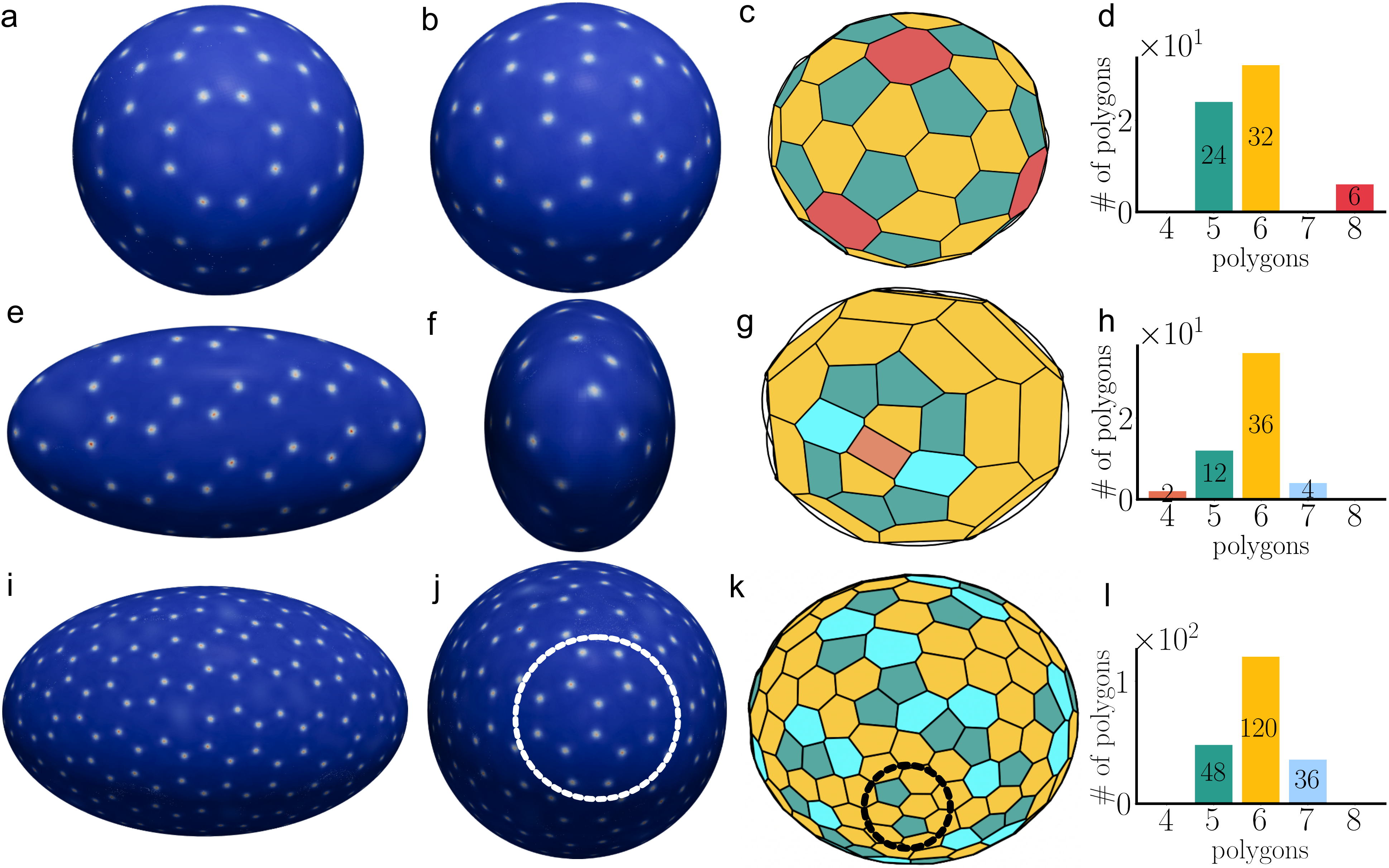}
\caption{\textbf{Ellipsoidal shells.} \textbf{(a-b)} Heat map of the biaxility parameter $c_p$ of an $OHP$ configuration ($R=50 , L=0.03$) from two different angle views. \textbf{(e-f)} Heat map of $c_p$ for a triaxial prolate ellipsoid (with semiaxes $a=41, b=82, c=30$) with approximately the same surface area of the $OHP$ configuration shown in the first row. \textbf{(i-j)} Heat map of $c_p$ for a triaxial ellipsoid with $a=80, b=70, c=120$ (same surface area of a speherical shell with $R=90$ in the amorphous phase).  
For each of these three configurations the corresponding Voronoi tessellation is reported in the third column (panels c, g, and k) where squares have been drawn in orange, pentagons in green, hexagons in yellow, heptagon in cyan and octagons in red. The histogram of the polygonal occurrences for each configuration is reported in the forth column in panels d,h, and l.}.
\label{fig5}
\end{figure*}

In Fig.~\ref{fig3}(a-f) we report  the probability density function  ($pdf$) of $\psi_n$ for  the equilibrium configurations of Fig.~\ref{fig1}(c-f) and Fig.~\ref{fig2}(b-c).\footnote{The distributions have been normalized in such a way that sum of the areas in each panel sums up to $1$, so that the area under the $\psi_n$ curve is a measure of the relative occurrence of a $n-$edged polygon in the corresponding tessellation.} As expected, for each quasi-crystalline configuration in panels (a-c) and (e), the non-zero distributions reflect the symmetries of the underlying polygonal tessellation. These are peaked on the predicted values (see figure caption and following discussion), signalling the local regularity expected for these phases. For instance, in the case of the $HP$ configuration each hexagon borders 3 pentagons and 3 hexagons in an alternated fashion, while pentagons only border hexagons. Therefore, the expected values for the order parameter $\psi_n$ are $\langle \psi_6 \rangle =(6+5)\pi/3=11.51$ and $\langle \psi_5 \rangle =2(6+6)\pi/5=15.07$, in excellent agreement with the measured distributions in panel (e). However, the distributions of quasi-crystalline structures tend to spread around their nominal value due to the slight deformation of the polygons with respect to their ideal shape. 

It is interesting to compare the $pdf$ for the $HP$ (Fig.~\ref{fig3}e) and the $HP^*$  (Fig.~\ref{fig3}f) phases, which shows the effect of the occurrence of heptagonal and pentagonal \emph{impurities}, effectively forming dislocations in the polygonal tessellation. Indeed, by looking at the $\psi_5$ distributions in the two cases, one can see that in the $HP^*$ case a tailed profile takes the place of the sharply peaked profile for the $HP$ configuration, due to the fact that the regular patterning of hexagons and pentagons is occasionally disrupted by the presence of a heptagon. We argue that this behavior is symptomatic of a continuous transition from the ordered to the amorphous phase occurring due to the proliferation of such dislocations~\cite{noi}, only occurring \emph{in absence} of tangential anchoring.
Finally, panel (d) shows the distributions for the amorphous configuration reported in Fig.~\ref{fig1}f. These are qualitatively different than those of the qusi-crystalline states commented above, as they spread out over a much wider range of values, thereby identifying these structures as amorphous.  

\subsection{Structure Factor}
\label{sec:struct_fact}
A further quantitative characterization of the topological phases we found is based on the computation of the  Bessel or spherical structure factor for the polygonal tessellation of each configuration.
This is done by expanding in spherical harmonics the spatial distribution of the polygon vertices~\cite{copar2019} (the $-1/2$ defects) determined via their polar coordinates $(\theta, \varphi)$. This distribution is encoded by the function
$$
g(\theta, \varphi) = 
\begin{cases}
1/\sqrt{V} \qquad &\text{at $-1/2$ defects' position} ,\\
0 \qquad &\text{otherwise},
\end{cases}
$$
where $V$ is the total number of defects in the tessellation.
The  Bessel expansion of $g(\theta, \varphi)$ consists in finding a set of suitable coefficients $\lbrace c_{lm} \rbrace$ such that
$$
g (\theta, \varphi) = \sum_{l=0}^{\infty} \sum_{m=-l}^{l} c_{lm} Y_{lm} (\theta, \varphi)
$$
where $Y_{lm}(\theta, \varphi)$ denote the spherical harmonic function of degree $l$ and order $m$ ($l=0,1,2,\dots$ and $m= -l, -l+1, \dots, l-1, l$). The estimate of the coefficients $\lbrace c_{lm} \rbrace$ is carried out numerically, by implementing the Python Spherical Harmonic Tools (SHTools)~\cite{shtools}.
With the coefficients $\lbrace c_{lm} \rbrace$ at hand, we next compute the spherical structure factor 
$$
S(l) = \sum_{m=-l}^{l} \dfrac{c_{lm}^2}{2l+1}
$$
and  normalize it  in accordance to the $4\pi$ rule   
$$
1 = \int d \Omega \ g(\theta, \varphi) = 4 \pi \sum_{l=0}^{\infty} \sum_{m=-l}^{l} c^2_{lm},
$$
where  $d \Omega$ is the solid angle element\footnote{Our algorithm was tested to give the same results as in~\cite{copar2019} for the solution of the Thomson problem for the optimal location of a generic number of charged particles on a sphere interacting via a Coulomb potential.}.

The results of the spectral analysis are shown in Fig.~\ref{fig4} for two quasi crystalline configurations --namely the $HP$ state in panel (a) and the $OHS$ state in panel (b)-- and for the amorphous configuration of Fig.~\ref{fig1}(c). 
For the two quasi crystalline cases, spectra exhibit a structure characterized by peaks, which are higher for low-order harmonics and progressively flatten at larger $l$. The $HP$ spectrum in panel (a) is also compared with the solution of the Thomson problem with 32 points\footnote{Topological phases other than $HP$ with more than $32$ defects do not exhibit the same symmetries as any other solution of the Thompson problem, despite their quasi-crystalline order.}. In this case the solution admits an exact geometrical solution which corresponds to a pentakis dodecahedron --the polyhedron constructed by attaching a pentagonal pyramid to a regular dodecahedron. This is the dual of the truncated icosahedron (a football) and therefore it has the same geometrical properties of the \emph{HP} configuration observed in simulations.
The spherical spectra of the $HP$ configuration and the corresponding Thomson problem are in good agreement and  they allow a determination of the leading harmonics in the structure. For both of them these correspond to angular momenta $l=10,18,36$. 

It is instructive to understand how each harmonic contributes to the final structure. 
To this aim, we consider the anti-Bessel transform of a superposition of only this minimal set of relevant harmonics. Panel (d) shows the carpeting obtained by means of the Mercator projection from the sphere onto the plane, while the sequence of panels (e)-(h) shows the superposition of various harmonics in Bessel expansion. The result clearly show that a minimal set of harmonics is enough to reconstruct the relevant features of the structure, while higher order harmonics serve to refine the structure. 
An analogous analysis for other finite quasicrystals (\emph{fqc}) configuration confirms that primary peaks in the spherical structure factor identify the harmonics which have the same geometrical structure of the observed configuration and they usually exhibit a certain periodicity. 
Conversely, in the case of the amorphous configuration the spectrum shown in panel~(c) hardly exhibits peaks and it is impossible to determine any geometrical symmetry based on the Bessel structure factor, as expected for an amorphous configuration.
We stress that, even in this case, a direct comparison of the structure factor computed for analogous configurations obtained with or without anchoring\cite{noi} does not exhibit significant quantitative differences.

\section{Curvature effects}
\label{sec:curvature}
In this Section we explore the role of non-homogeneous curvature in the formation and placement of defects on the surface of the shell.
To this aim we will first consider ellipsoidal shells,  that are topologically equivalent to a sphere (as both have genus $0$, or no holes), but have non-constant curvature. Next, we consider the case of toroidal shells, which are topologically not equivalent to a sphere (as they have genus $1$, or one hole), and which possess coexisting phases localised within regions of positive and negative curvature.

\begin{figure}[]	\centering\includegraphics[width=1.0\columnwidth]{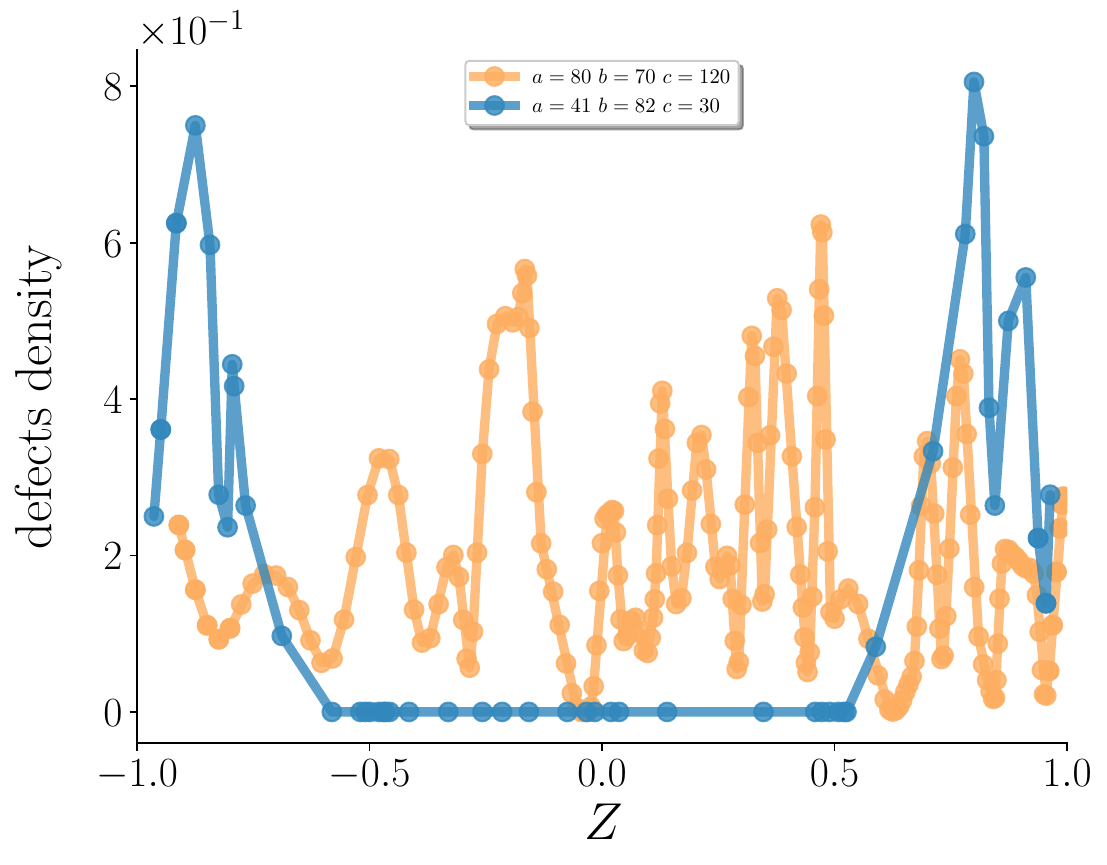}
	\caption{\textbf{Curvature effects on defects density.} Profile of the defect density of the tessellation (see text) for the ellipsoids in Fig.~\ref{fig5}(e-f) (blue) and (i-l) (orange) as a function of the position along the long axis ($Z$ axis) of the ellipsoid. Both the defect density and the $Z$-position are normalized with respect to the total defect charge and the semi-axis of the ellipsoid.}
	\label{fig6}
\end{figure}

\begin{figure*}[t!]
	\centering\includegraphics[width=1.9\columnwidth]{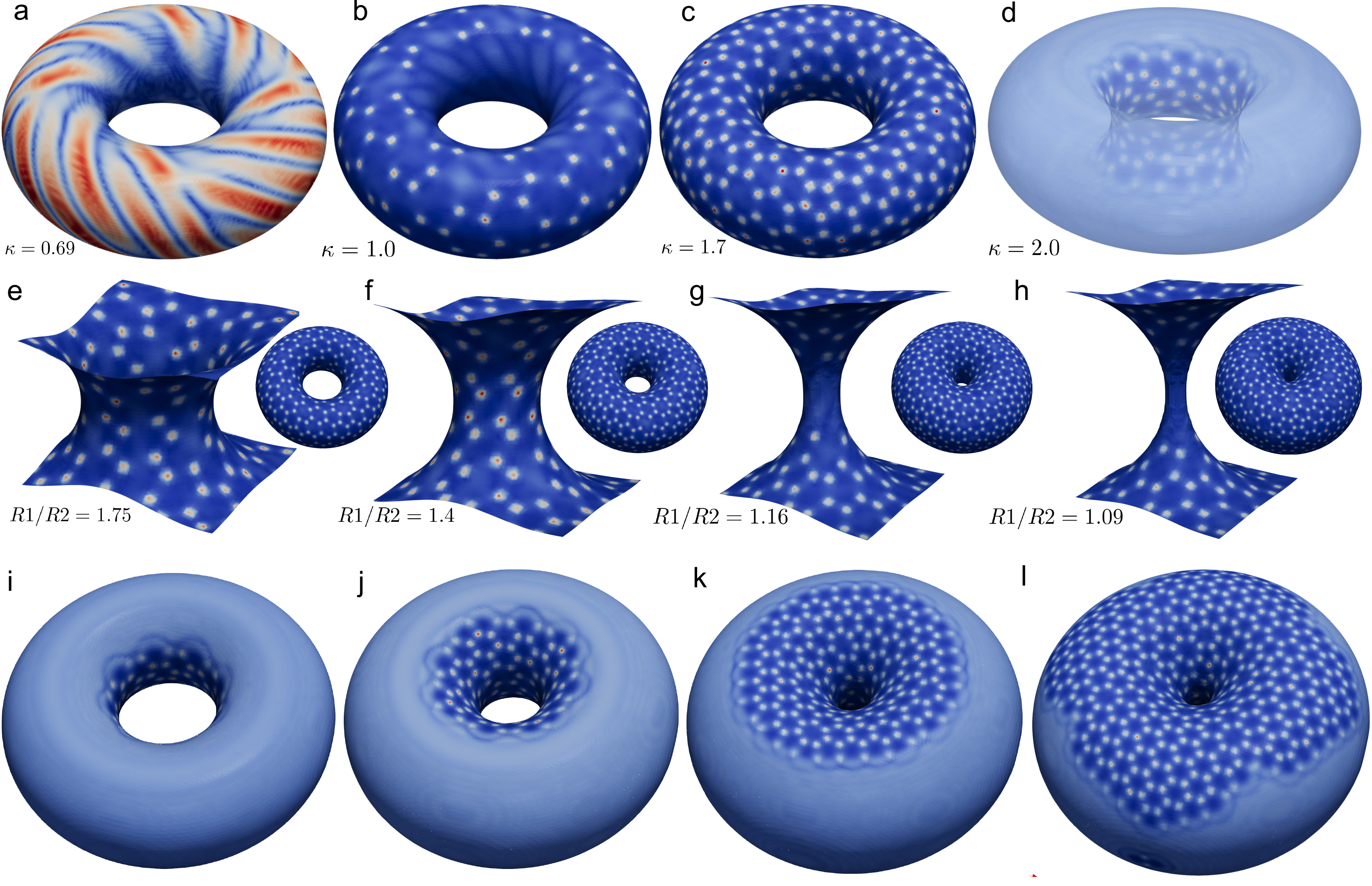}
	\caption{\textbf{Toroidal shells.} \textcolor{black}{\textbf{(a)-(d)} Patterns  of the biaxiality parameter $c_p$  on toroidal shells with $R1/R2=2$ for different chirality strength $\kappa$. In the cholesteric phase of panel~(a) at $\kappa=0.69$, the director field winds around the torus -- color code as in Fig.~\ref{fig1}. At $\kappa=1.0$ (panel~(b)) a heterogeneous pattern emerges with a lattice of topological defects (red spots) only in the regions with positive curvature. At $\kappa=1.7$ the lattice occupies the whole surface. At $\kappa=2.0$ the systems undergoes a cholesteric-isotropic transition, with defects only present in the regions with negative curvature, while the isotropic phase covers the positive curvature regions.
	\textbf{(e)-(h)} Patterns  of the biaxiality parameter $c_p$  on the negative curvature region of toroidal shells with chirality strength $\kappa=1.7$, obtained by varying $R1/R2$. 
	\textbf{(i)-(l)} Patterns  of the biaxiality parameter $c_p$ on the surface of  toroidal shells with chirality strength $\kappa=2.0$, for the same values of $R1/R2$ as in panels (e)-(h).} }
	\label{fig7}
\end{figure*}

\subsection{Ellipsoidal shells}
Shells with ellipsoidal shape are obtained by choosing the $\mathcal{T}(\bm{r})$ function in Eq.~\eqref{eqn:parametric_non_spherical} as the parametric equation of an ellipsoid, and initializing an ellipsoidal droplet with $\phi=\phi_0$ in those regions where $\mathcal{T}(\bm{r})<0$ and $0$ otherwise. Then, we let the $\phi$ field thermalise with $\bm{Q}=\bm{I}$ until the interface is fully formed. Once no significant variation is observed in the time evolution of $F^\phi$, we treat $\phi$ as a static field and let the Q-tensor relax as previously done for the case of spherical shells\footnote{An analogous procedure was carried out to simulate toroidal shells, to be discussed in the following section, with $\mathcal{T}(\bm{r})=0$ defining the parametric equation of a torus.}.

With the aim of understanding the role of curvature in the patterning of defects, we now fix the cholesteric strength to $\kappa = 0.78$ and we consider ellipsoids with different shapes and sizes.
The Gaussian curvature can be calculated through the gradients of the vector $\bm{m(r)}$ which locally defines the normal to the manifold as 
$$
\kappa_G(\mathbf{r}) = \dfrac{1}{2} \nabla \cdot \left[ (\bm{m(r)}\cdot \nabla) \bm{m(r)} - (\nabla \cdot \bm{m(r)} ) \bm{m(r)} \right] .
$$ 
In our approach the normal vector $\bm{m(r)}$ can be easily computed as the normalized gradient of the $\phi(\bm r)$ field, namely $\bm{m(r)} = \nabla \phi(\bm r) / |\nabla \phi(\bm r)|$.
We start by comparing the $OHP$ configuration for a spherical shell, (Fig.~\ref{fig5}a-d) with that obtained for a triaxial prolate ellipsoid having approximately the same surface (panels e-h), so that the total number of polygons in the tessellation is comparable ($62$ for the sphere and $54$ for the ellipsoid).
We chose the axis of the ellipsoid (see caption for values) so as to create a large variation in the Gaussian curvature $k_G$ ($\max(k_G)/\min(k_G)=55.8$)
Interestingly, the regular pattern of the \emph{fqc} configuration on the sphere is lost for the ellipsoid, with a large stack of hexagons carrying no topological charge in the tessellation inhabiting the central part of the ellipsoid.
This new distribution of defects still satisfies the Euler equation $\sum \left( 1 - \frac{n}{6}\right )=2$ for a spherical topology, but now the defects with larger positive charge (the squares, $s =1/3$) are placed at the antipodes of the ellipsoid, i.e. the regions  with the  highest positive curvature, and are surrounded by two heptagons ($-1/3$ total charge) and 6 pentagons ($+1$ total charge). 
In general, the population density of  defects is not homogeneously distributed on surfaces with non-homogeneous Gaussian curvature and it is larger in proximity of the region of higher curvature. To quantitatively substantiate this statement, we consider the distribution of defective polygons (with a number of edges different than $6$). We plot their number density profiled along the long axis of the ellipsoid in Fig.~\ref{fig6} (blue points), clearly showing the accumulation of defects at the extremes ($|Z| \simeq 1$) where the curvature is larger~\footnote{We consider as \emph{defects} in the tessellation any polygon with a number of edges other than $6$. The profile in Fig.~\ref{fig6} were obtained by slicing the ellipsoidal shells and computing the number density of defective polygons in each slice.}.

We now consider the case of a larger prolate ellipsoid (Fig.~\ref{fig5}i-l) whose total surface corresponds to a spherical shell with $R/p_0\sim 3.12$ in the amorphous phase. In this case, the curvature is less heterogeneous ($\max(k_G)/\min(k_G)=8.6$) and its effect on the defect arrangement less critical. We observe instead the proliferation of a large number of tightly bound $5-7$ defects (as each $5-7$ pair has a net total topological charge of $0$) that are organised as chains called \emph{scars} in the literature~\cite{Irvine2010}, in such a way that the observed configuration is akin to an $HP^*$ configuration, with elongated scars of pentagon-heptagon pairs spanning the whole shell (panel~k).
Interestingly, we notice that pentagons whose charge is not locally neutralised by the presence of a contacting negative charge heptagons are placed in proximity of the regions with the largest positive curvature (dashed circles in Fig.~\ref{fig5}j-k).
This behaviour is confirmed by the profile of the number density of non-hexagonal polygons in Fig.~\ref{fig6} (orange line) which shows that, in this case, defects populate the entire ellipsoidal surface, regardless of curvature.

We argue that there are two main physical effects underlying these results. First,  defects with the largest positive topological charges are attracted by regions of higher positive curvature. When this effect is dominant, for instance in smaller shells with large curvature variations, all the topological charge can be localized in region of maximal curvature, while the rest of the shell is left free of defects.
Analogous behaviors were previously reported  in the literature in systems such as elastic curved crystals~\cite{Jimenez_et_al_PRL_2016} and emulsions of two immiscible fluids stabilised by a packed film of colloidal spheres at the droplet fluid interface~\cite{Burke_et_al_Soft_Matter_2015}. Interestingly, this effect also appears to play an important role in animal morphogenesis --the process whereby animal tissues attain their complex shapes-- as recently discussed in~\cite{MetselaarHydra,maroudasack,livio_science_advances}.

Second, larger system, for which geometrical and topological confinement is less relevant, can relieve large strains associated with quasi-crystalline structures by forming large scars of dislocation pairs with null topological charge, consistently with our previous discussion on the origin of the $HP^*$ phase in Section~\ref{sec:phasediagram}, even for non-spherical geometries.

\subsection{Toroidal shells}
The reason to look at the formation and stability of topological phases on toroidal shells is two-fold. First, it is the simplest example of  a surface with Euler characteristic different from the spherical one ($s=0$ for toroidal surfaces while $s=2$ for spheres). Second, unlike spheroids, toroidal surfaces display also saddle-like regions of negative Gaussian curvature. Having regions with both positive and negative Gaussian curvature suggests that the final defect structures could be richer than those observed on spheroids where only the chirality $\kappa$ and the extension of the confining geometry determine the final equilibrium configuration.  For instance, in the case of nematic order, it is known that $+1/2$ and $-1/2$ defects are attracted by regions of Gaussian curvature of the same sign as their topological charge, causing the unbinding of a defect dipole of total charge zero~\cite{PhysRevE.69, giomiepje,C4SM02540G}. 

To explore the effect of non-constant positive/negative curvature and cholesteric strength $\kappa$ we first focus on toroidal shell with radii $R_1=36$ and $R_2=18$ ($R_1/R_2=2$) corresponding to an area extension comparable to that of a spherical shell with radius $R=44$. Since $q_0=0.245$ this corresponds to $R/p_0\approx 1.71$. By varying $\kappa$ from $0.7$ to $2.0$ and letting the system to thermalize  various phases are detected; these are reported in Fig.\ref{fig7}(a-d).
The first thing to note is that the presence of a region of negative curvature modifies the phase diagram previously obtained for the spherical shells with the same surface extension (see Fig.~\ref{fig1}(a)). More precisely, at $\kappa=0.7$ (corresponding to the $OHS$ phase for the spherical case)  we observe here a defect-free cholesteric phase, characterized by a helical pattern in the director field winding all around the torus in a continuous fashion.
Increasing chirality strength to $\kappa=1.0$ (panel (b)), topological defects of charge $-1/2$, delimiting half-skyrmions structures, appear on the regions of positive curvature while in the saddle regions, where the curvature is negative, the arrangement remains helical and defect-free. It is interesting to notice that the non-trivial geometry of the confining manifold can directly affect and spatially modulate the transition between topological phases. 

\begin{figure}[t!]
\center
\includegraphics[width=1.0\columnwidth]{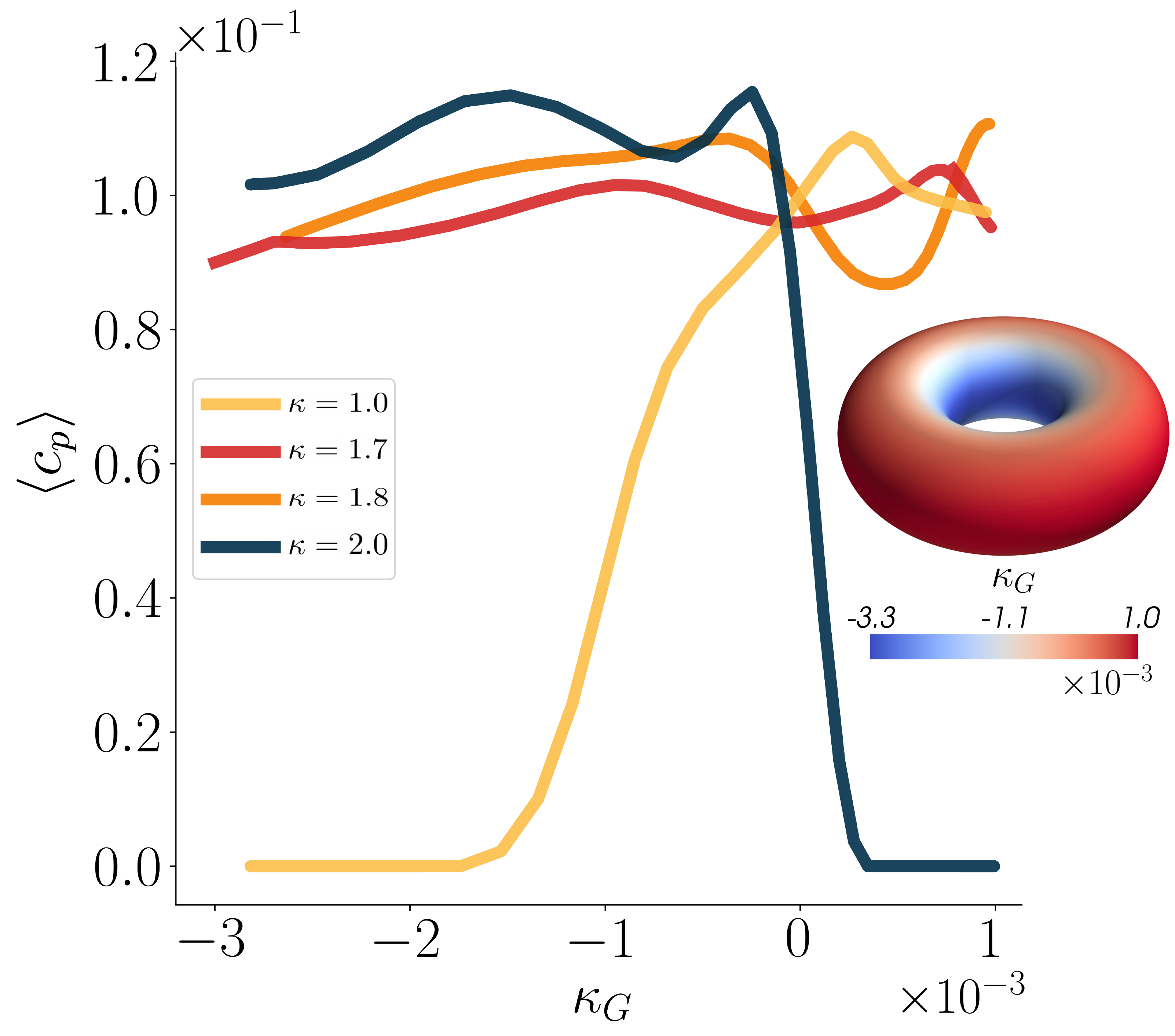}
\caption{Average biaxility parameter  $\langle c_p \rangle$ against the curvature $k_G$, for the cases shown in panels (b-d) of Fig.\ref{fig7}. Inset shows the colour plot of the Gaussian curvature $k_G$.}
	\label{fig8}
\end{figure}

For $\kappa=1.7$ (panel (c)) the half-skyrmion lattice  covers the entire surface of the torus while, at sufficiently large values of $\kappa$ (see panel (d) for $\kappa=2.0$), there is an inversion of the phase coexistence, where  half-skyrmions concentrate in the negative curvature regions  while in the positive ones, the phase becomes isotropic. This inversion  is more apparent if one profiles the value  of the biaxiality parameter $c_p$ against the Gaussian curvature of the surface. Fig.~\ref{fig8} Shows the average of the biaxiality parameter $\langle c_p \rangle$,  across circles of different Gaussian curvature $k_G$, in the toroidal direction~\footnote{We notice that the biaxiality parameter offers an excellent quantitative tool to estimate not only the biaxiality of the liquid crystal, but also the position of topological defects. Regions characterized by a larger average value of the biaxiality parameter are therefore, also richer of disclinations.}.
These results suggest the existence of two transitions for cholesteric toroidal shells, upon changing chirality $\kappa$ and whilst keeping $R_1/R_2$ constant. There is first a transition from cholesteric to a half-skyrmion lattice that covers the entire shell surface, with a coexistence region between the two phases, and a second transition to the isotropic phase for larger values of $\kappa$, again with a region of coexistence. This behaviour is not observed for spherical shells, and is an effect of the non-constant positive/negative curvature of the confining toroidal geometry.

To further investigate the coexistence between different topological phases due to non-uniform curvature, we now fix the value of the reduced chirality ($\kappa=1.7$) and vary the minor radius of the torus $R_2$ from $20$, as in panel~(c), to $30$. This corresponds to a variation of the ratio $R_1/R_2$ from $1.7$ to $1.09$ which allows us to explore regions with increasing negative curvature in the saddle region.
The results of this analysis are graphically shown in panels (e-h) where the torus core has been separately plotted to ease visualization:  regions with larger negative curvature, panels (g-h), can host helicoidal phase, replicating the phase-coexistence that was already observed for the case at lower chirality ($\kappa=1.0$) shown in panel (b).
This finding suggests that chirality and curvature could be both used to regulate the transition between different topological phases and, in more detail, that increasing negative curvature can mimic the effect of reducing the chirality of the liquid crystal.

To substantiate this hypothesis, we perform a similar experiment starting from the case at $\kappa=2.0$, panel (d), where coexistence between different topological phases was observed --namely half-skyrmion tessellation and isotropic phase-- and decrease the ratio $R_1/R_2$ to check weather we can extend the half-skyrmion phase by varying the curvature. The configurations shown in panels (i)-(l) confirm that this is the case: as larger $R_2$ are considered, a lattice of $-1/2$ defects invades a progressively larger area fraction of the toroidal shell.

\section{Conclusions}
\label{sec:conclusions}
To conclude, we investigated the nature of the topological phases arising in non-Euclidean cholesteric shells close to the isotropic-cholesteric transition. We have shown that the curved geometry frustrates the formation of regular half-skyrmion lattices, which are instead stable on flat surfaces, due to topological constraints arising from the Gausee-Bonnet theorem.

On a spherical shell, for intermediate chirality and small radii the emerging structures are \emph{fqc} composed by a network of surface defects with topological charge $-1/2$. These structures can be seen as polyhedra composed of regular polygons, corresponding to half-skyrmion tessellations of the surface of the sphere. 
For larger shells, a qualitatively distinct amorphous phase develops: this is characterized by a disordered arrangement of polygons on the shell, reminiscent of the three-dimensional structure of blue phase III, which can be viewed as an amorphous network of disclination lines~\cite{Henrich2011}. Simulations suggest that, like blue phase III,  the amorphous phase is thermodynamically stable in a finite parameter range. Interestingly, we have shown that orientational anchoring of the director field at the interface, favouring for instance tangential, or planar, alignment, significantly affects the topological transition between quasicrystalline and amorphous tessellation. The transition occurs directly in the case with anchoring, whereas in the free anchoring case there is an intermediate regime characterised by the nucleation of dislocation scars --namely, chains of heptagons and pentagons with zero topological charge-- similarly to what happens in spherical crystals~\cite{Bowick2000}.

After investigating the effects of topological confinement, we switched our attention to geometrical effects, \emph{i.e.} we studied how non-uniform curvature affects the stability of topological phases. Interestingly, we found that polygons with a number of edges less than $6$ are attracted towards areas with larger Gaussian curvature. We argue that this effect can be rationalized by interpreting $n$-edges polygons in the tessellation as defective structures carrying a topological charge $q=1-n/6$ so that these are attracted by positive (negative) Gaussian curvature for $n<6$ ($n>6$) to minimize elastic deformations of crystalline structures.
In this respect, our findings pose the intriguing question of how external perturbations which are able to change the shell shape --a shear flow, for instance-- might affect the relaxation properties of quasi-crystalline configurations. We believe that addressing this question will be relevant to understanding how to use external cues and geometry to self assemble a desired defect configuration.

The pathway described above is not the only one through which geometry affects pattern formation in our system. Indeed, we showed that non-uniform curvature can be exploited to stabilise coexistence between different topological phases. In particular, we considered the case of toroidal shells which feature regions with both positive and negative Gaussian curvature. At intermediate chirality we observed coexistence between a helical phase and a hexagonal half-skyrmion tessellation, respectively localised in regions of negative and positive curvature. At larger chirality and close to the isotropic transition, coexistence between isotropic and hexagonal phase was instead observed. The selection of which phases appear is determined by a competition between chirality, which favours the isotropic phase, and curvature, which effectively renormalizes the chirality and can stabilize hexagonal patterning even in regions that would be otherwise isotropic. These considerations gain even more relevance by observing that recent experimental studies \cite{Norouzi2022,Tran2020,Durey2022, Laventrovich2020, Tran2018} on cholesteric shells reported that homeotropic anchoring can result in fingering patterns or arrays of skyrmions, and found that the different topological phases where actually controllable by applying an alternating electric field or changing the shell thickness, a case of study which we will address in future work. 

Besides being of theoretical interest to understanding liquid crystalline ordering and topological phases on non-Euclidean geometries, we are positive that our work will stimulate future experimental research in the field of cholesteric shells. Nowadays it is now possible to self-assemble these systems in the lab, for instance by confining liquid crystals to the surface of emulsion droplets; toroidal geometries can also be created by shaping oil droplets in shear-yielding materials. Such realisations could be used to test our numerical predictions experimentally.

\section*{Acknowledgements}
This work was carried out on the Dutch national e-infrastructure with the support of SURF through Grant 2021.028 for computational time (L.N.C and G.N.).  We acknowledge funding from MIUR Project No. PRIN 2020/PFCXPE.
\balance

\appendix
\label{sec:appendix}
\section{Numerical Method and parameters}
To solve the hydrodynamics of the system, we made use of a hybrid lattice Boltzmann (LB) approach on a $D3Q15$ lattice~\cite{succi1991,succi2018}. The Navier-Stokes equation was solved through a predictor-corrector LB scheme~\cite{denniston2001,carenzareview}, while the evolution equations for the order parameters $\phi$ and $\mathbf{Q}$ were integrated through a predictor-corrector finite-difference algorithm implementing first-order upwind scheme and fourth-order accurate stencils for space derivatives.
Within this approach the evolution of the fluid is described in terms of a set of distribution functions ${f_i(\mathbf{r}_\alpha,t)}$ (with the index $i$ labelling different lattice directions, ranging from $1$ to $15$ for our D3Q15 model) defined on each lattice site $\textbf{r}_\alpha$. Their evolution follows a discretised predictor-corrector version of the Boltzmann equation in the Bhatnagar-Gross-Krook (BGK) approximation:
\begin{multline}
f_i (\mathbf{r}_\alpha + \bm{\xi}_i \Delta t) - f_i (\mathbf{r}_\alpha,t) = \\ - \dfrac{\Delta t }{2} \left[ \mathcal{C}(f_i,\mathbf{r}_\alpha, t) + \mathcal{C}(f_i^*,\mathbf{r}_\alpha+ \mathbf{\xi}_i \Delta t, t) \right].
\label{eqn:LBevolution}
\end{multline} 
Here $\lbrace \bm{\xi}_i \rbrace$ is the set of discrete velocities of the $D3Q15$ lattice.
Here, $\mathcal{C}(f,\mathbf{r}_\alpha, t)=-(f_i-f_i^{eq})/\tau + F_i$ is the collisional operator in the BGK approximation, expressed in terms of the equilibrium distribution functions $f_i^{eq}$ and supplemented with an extra forcing term for the treatment of the anti-symmetric part of the stress tensor.
The distribution functions $f_i^*$ are first-order estimations to  $f_i (\mathbf{r}_\alpha + \bm{\xi}_i \Delta t) $ obtained by setting $f_i^* \equiv f_i$ in Eq.~\eqref{eqn:LBevolution}. 
The density and momentum of the fluid are defined in terms of the distribution functions as follows:
\begin{equation}
\sum_i f_i = \rho \qquad \sum_i f_i \bm{\xi}_i = \rho \mathbf{v}.
\label{eqn:variables_hydro}
\end{equation}
The same relations hold for the equilibrium distribution functions, thus ensuring mass and momentum conservation. 
In order to correctly reproduce the Navier-Stokes equation, the following conditions on the second moment of the equilibrium distribution functions are imposed:
\begin{equation}
\sum_i f_i \bm{\xi}_i \otimes \bm{\xi}_i = \rho \mathbf{v} \otimes \mathbf{v} -\tilde{\sigma}^{el}_s,
\label{eqn:constrain_second_moment}
\end{equation}
whilst the following condition is imposed for the force term:
\begin{equation}
\sum_i F_i = 0, \qquad \sum_i F_i \bm{\xi}_i = \mathbf{\nabla} \cdot \tilde{\sigma}^{el}_a, \qquad \sum_i F_i \bm{\xi}_i \otimes \bm{\xi}_i = 0.
\label{eqn:constraint_force}
\end{equation}
In the equations above, we respectively denoted with $\tilde{\sigma}^{el}_s$ and $\tilde{\sigma}^{el}_a$ the symmetric and anti-symmetric part of the elastic stress tensor.
The equilibrium distribution functions are expanded up to second order in the velocities, as follows:
\begin{multline}
f_i^{eq} = A_i + B_i (\bm{\xi}_i \cdot \mathbf{v}) + C_i |\mathbf{v} |^2 + \\ D_i (\bm{\xi}_i \cdot \mathbf{v})^2 + \tilde{G}_i : (\bm{\xi}_i \otimes \bm{\xi}_i).
\end{multline}
Here cthe oefficients $A_i, B_i,C_i,D_i,\tilde{G}_i$ are to be determined by imposing the conditions in Eqs.~\eqref{eqn:variables_hydro} and \eqref{eqn:constrain_second_moment}. In the continuum limit the Navier-Stokes equation is restored if $\eta=\rho\tau/3$~\cite{denniston2001}.

We performed simulations on 3D square lattices of size $L$ ranging from $128$ to $384$.
Periodic boundary conditions were imposed at the boundary. 

Except otherwise stated, the system is initialized as follows:
\begin{itemize}
    \item spherical shell of radius $R$: the phase field $\phi(\bm{r})=\bar{\phi_0}=2.0$ for $|\bm{r}|<R$ inside the sphere, while $\phi(\bm{r})=0$ for $|\bm{r}|>R$ outside the sphere. For all cases shown in the main text, the Q-tensor is randomly initialized and $\bm{Q}^2 \ll 1$. 
    \item torus/ellipsoid: the phase field is set to $\bar{\phi_0}=2.0$ inside the torus/ellipsoid and null outside. Even in this case, the Q-tensor is randomly initialized with $\bm{Q}^2 \ll 1$.
\end{itemize}

Lattice units (LU) have been used to present result in the main text. The values of the model parameters in LU for the case of spherical droplets are $A_0=1.2 \times 10^{-1}$, $q_0=2.45 \times 10^{-1}$ (corresponding to an equilibrium pitch $p_0=2\pi/q_0 \simeq 25$), $a=10^{-2}$, $k_\phi=0.14$, $M=10^{-1}$, $\Gamma=1$, $\tilde{\xi}=0.7$. 
The radii of the droplets considered range from $R=25$ to $R=90$, while the elastic constant of the LC varies from $L=10^{-3}$ to $L=7 \times 10^{-2}$.
Two key control parameters in the model are: \emph{(i)} the reduced temperature $\tau={9(3-\chi)}/{\chi}$, and \emph{(ii)} the chirality strength $\kappa=\sqrt{108 q_0^2 L /(A_0 \chi)}$, proportional to the ratio between nematic coherence length and cholesteric pitch~\cite{alexander2009,Wright89}. In all simulations presented in the main text the reduced temperature is fixed to $\tau=0.540$, while the chirality varies from $\kappa=0.1$ to $\kappa=1.1$.  
Parameters are chosen in such a way to ensure that the droplet does not deform during the relaxation dynamics.
As in blue phases formed in cholesteric LCs in 3D, the free energy is expected to quantitatively depend in a sensitive way to changes in chirality and reduced temperature, as well as to variations in the elastic constants going beyond the one elastic constant approximation~\cite{Alexander2006}. The qualitative behaviour of the system is instead expected to be robust to such changes, and similar to the one reported in the main text.

A mapping from lattice units to physical units can be obtained by fixing the relevant length, time and force scales respectively given by $l^* = 10 n\text{m}, t^* = 1 m\text{s}, f^* = 1 \mu \text{N}$ which  are set to $1$ in LU\cite{negro2019}. Following Ref.~\cite{Popov2017}, for the case of 5CB liquid crystal with $5\%$ CD1 chiral dopant at $30 ^{\circ}$C we propose here an approximate estimation of this experimental setting in terms of our rescaled dimensionless units. Assuming for the elastic constant $L \approx 10$pN the average value of the three elastic contributions (splay, twist and bend) and a coherence length $\xi \approx 10$nm (see also Ref.~\cite{stark2001}), we can therefore estimate the value of the bulk constant $A_0$ from the relation $\xi=\sqrt{L/A_0}$. This gives for the reduce chirality $\kappa \approx 0.5$, while the radius of the confining shell for which we expect topological phases to be observed ranges between a few micrometers and $\approx 100 \mu$m. 

For the case of a torus the same set of parameters has been used, except for the elasticity of the LC which is fixed at $L=1.5 \times 10^{-2}$. LC chirality is changed by varying $q_0$ in the range $0 - 5.4 \times 10^{-1}$. For all cases discussed in the main text the major radius is $R_1=35$ and the minor radius $R_2=18$.

The degree of biaxiality of the LC has been computed by following the approach of Ref.~\cite{callan2006} as the second parameter of the Westin metrics $c_p = 2(\tilde\lambda_2-\tilde\lambda_3)$, where $\tilde\lambda_1$, $\tilde\lambda_2$ and $\tilde\lambda_3$ (with $\tilde\lambda_1\ge\tilde\lambda_2\ge\tilde\lambda_3$) are three eigenvalues of the positive definite matrix $G_{\alpha\beta}=Q_{\alpha\beta}+\delta_{\alpha\beta}/3$.

\bibliography{rsc} 

\end{document}